\documentclass{phb-proc4-auth}

\usepackage{graphicx}
\usepackage{amssymb}

\begin{document}
\begin{frontmatter}


\journal{SCES '04}


\title{%
Key Role of Orbital Anisotropy in Geometrically Frustrated Electron System}


\author{Hiroaki Onishi\corauthref{1}} and
\author{Takashi Hotta}


\address{%
Advanced Science Research Center,
Japan Atomic Energy Research Institute,
Tokai, Ibaraki 319-1195, Japan}




\corauth[1]{%
Corresponding Author:
Advanced Science Research Center,
Japan Atomic Energy Research Institute,
Tokai, Ibaraki 319-1195, Japan.
Email: onishi@season.tokai.jaeri.go.jp}


\begin{abstract}

By using the density matrix renormalization group method,
we investigate ground- and excited-state properties of
the $e_{\rm g}$-orbital degenerate Hubbard model at quarter filling
for two kinds of lattices, zigzag chain and ladder.
In the zigzag chain,
the system is effectively regarded as a decoupled double chain of
the $S$=1/2 antiferromagnetic Heisenberg model,
and the spin gap is approximately zero,
similar to the case of weakly coupled Heisenberg chains.
On the other hand, in the ladder,
the spin correlation on the rung remains robust
and the spin gap exists.

\end{abstract}




\end{frontmatter}


Frustration among competing interactions is a key issue
to bring richness of cooperative phenomena
in magnetic systems \cite{frustration-review}.
For instance,
antiferromagnets with the triangle-base structure possess frustration,
since it is impossible to satisfy all the interactions.
To suppress the effect of frustration,
spins compromise with each other to take
the noncollinear 120$^\circ$ structure.
On the other hand,
when electrons partially fill degenerate orbitals,
it is necessary to consider the orbital
as an internal degree of freedom in addition to the spin.
The spatial anisotropy of the orbital causes
electron hopping between orbitals depending on the direction,
hence it is important to take account of the lattice geometry
to understand the spin-orbital structure
even in a simple cubic lattice.
In particular, it is interesting to consider the effect of
the orbital anisotropy on the frustrated lattice.

\begin{figure}[t]
\centering
\includegraphics[width=\linewidth]{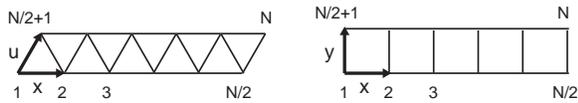}
\caption{\label{fig-lattice}%
Lattice configurations and numbering of the site.
The zigzag chain and the ladder are considered as two-chain systems
with the zigzag path and the rung, respectively.}
\end{figure}

In this paper, we investigate
the $e_{\rm g}$-orbital degenerate Hubbard model at quarter filling
on the zigzag chain and the ladder.
When the Hund's rule coupling $J$ is small,
it is found that the ground state is a paramagnetic (PM) phase
with ferro-orbital (FO) ordering.
Here we focus on the PM phase and set $J$=0 for simplicity.
The effect of $J$ will be discussed elsewhere in the future.
Then, the Hamiltonian is given by
\begin{eqnarray}
 H &=&
 \sum_{{\bf i},{\bf a},\gamma,\gamma^\prime,\sigma}
 t_{\gamma\gamma^\prime}^{\bf a}
 d_{{\bf i}\gamma\sigma}^{\dag} d_{{\bf i}+{\bf a}\gamma^\prime\sigma}
 \nonumber\\
 &&+
 U \sum_{{\bf i},\gamma}
 \rho_{{\bf i}\gamma\uparrow} \rho_{{\bf i}\gamma\downarrow}
 +U^\prime \sum_{{\bf i},\sigma,\sigma^\prime} 
 \rho_{{\bf i}a\sigma} \rho_{{\bf i}b\sigma^\prime},
\end{eqnarray}
where $d_{{\bf i}a\sigma}$ ($d_{{\bf i}b\sigma}$)
is the annihilation operator for an electron with the spin $\sigma$
in the $d_{3z^2-r^2}$ ($d_{x^2-y^2}$) orbital at the site ${\bf i}$,
$\rho_{{\bf i}\gamma\sigma}$=%
$d_{{\bf i}\gamma\sigma}^{\dag}d_{{\bf i}\gamma\sigma}$,
and ${\bf a}$ is the vector connecting adjacent sites.
Note that the lattices are arranged in the ($x,y$)-plane
as shown in Fig.~\ref{fig-lattice}.
The hopping amplitudes are given by
$t_{aa}^{\bf x}$=$t/4$,
$t_{ab}^{\bf x}$=$t_{ba}^{\bf x}$=$-\sqrt{3}t/4$,
$t_{bb}^{\bf x}$=$3t/4$ for the $x$-direction,
$t_{aa}^{\bf y}$=$t/4$,
$t_{ab}^{\bf y}$=$t_{ba}^{\bf y}$=$\sqrt{3}t/4$,
$t_{bb}^{\bf y}$=$3t/4$ for the $y$-direction, and
$t_{aa}^{\bf u}$=$t/4$,
$t_{ab}^{\bf u}$=$t_{ba}^{\bf u}$=$\sqrt{3}t/8$,
$t_{bb}^{\bf u}$=$3t/16$ along the zigzag path.
Hereafter, $t$ is taken as the energy unit.
We employ the density matrix renormalization group method
with the finite-system algorithm \cite{White-DMRG},
which allows us to investigate the frustrated system with high accuracy.
Here we consider each orbital as a virtual site
to reduce the size of the superblock Hilbelt space.
The number of states kept for each block $m$ is up to $m$=200
and the truncation error is estimated to be 10$^{-5}$ at most.

\begin{figure}[t]
\centering
\includegraphics[width=\linewidth]{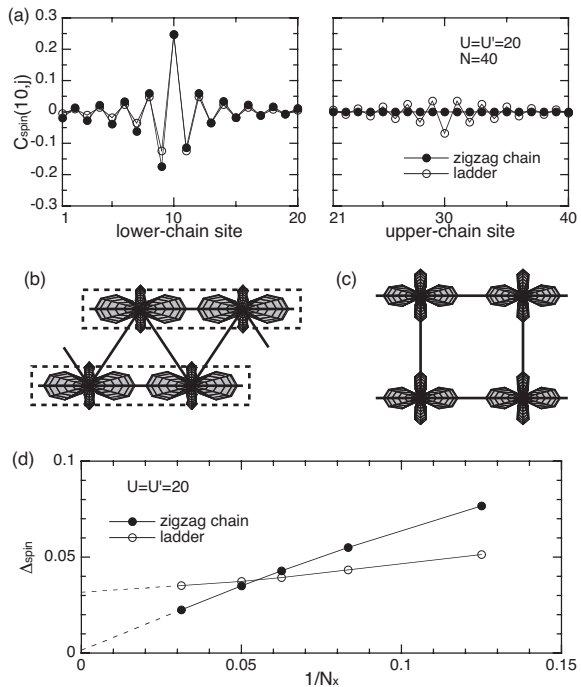}
\caption{\label{fig-result}%
(a) The spin correlation functions for the lower- and upper-chains
from the center of the lower chain in the PM ground state.
The FO structures for (b) the zigzag chain and (c) the ladder.
(d) The spin gap vs. 1/$N_x$ with $N_x$=$N$/2.
The dotted line denotes the linear extrapolation to $N_x$=$\infty$.}
\end{figure}

In the strong-coupling region,
each site is occupied by one electron, indicating that
only spin and orbital degrees of freedom remain effective.
In Fig.~\ref{fig-result}(a),
the spin correlation function
$C_{\rm spin}({\bf i},{\bf j})$=$\langle S_{\bf i}^z S_{\bf j}^z \rangle$
with
$S_{\bf i}^z$=
$\sum_{\gamma}(\rho_{{\bf i}\gamma\uparrow}$$-$%
$\rho_{{\bf i}\gamma\downarrow})/2$
is shown for $U$=$U^\prime$=20.
In the zigzag chain,
we find that the antiferromagnetic (AFM) correlation exists
between intra-chain sites in each of a double chain,
while the spin correlation between inter-chain sites is much weak.
The double-chain spin structure is understood
from the FO structure.
In order to characterize the orbital shape,
it is useful to define an angle $\theta_{\bf i}$ at each site as
\begin{equation}
 \tan\theta_{\bf i}=
 \langle T_{\bf i}^x \rangle / \langle T_{\bf i}^z \rangle
\end{equation}
in the spirit of the mean-field approximation \cite{Hotta-manganites},
where
$T_{\bf i}^x$=%
$\sum_{\sigma}
(d_{{\bf i}a\sigma}^{\dag}d_{{\bf i}b\sigma}$+%
$d_{{\bf i}b\sigma}^{\dag}d_{{\bf i}a\sigma})/2$
and
$T_{\bf i}^z$=%
$\sum_{\sigma}
(d_{{\bf i}a\sigma}^{\dag}d_{{\bf i}a\sigma}$$-$
$d_{{\bf i}b\sigma}^{\dag}d_{{\bf i}b\sigma})/2$.
By using $\theta_{\bf i}$,
the new operators are defined as
$\tilde{d}_{{\bf i}a\sigma}$=%
$e^{i\theta_{\bf i}/2}[\cos(\theta_{\bf i}/2)d_{{\bf i}a\sigma}$+%
$\sin(\theta_{\bf i}/2)d_{{\bf i}b\sigma}]$
and
$\tilde{d}_{{\bf i}b\sigma}$=
$e^{i\theta_{\bf i}/2}[-\sin(\theta_{\bf i}/2)d_{{\bf i}a\sigma}$+%
$\cos(\theta_{\bf i}/2)d_{{\bf i}b\sigma}]$
\cite{Hotta-berryphase},
which determine the orbital shape.
For the zigzag chain,
$\theta_{\bf i}$$\simeq$$4\pi/3$ is observed,
indicating the (3$x^2$$-$$r^2$)-type FO structure
shown in Fig.~\ref{fig-result}(b).
The orbital shape extends along the double chain,
not along the zigzag path,
to gain the kinetic energy by electron hopping.
Namely, the AFM superexchange interaction along the $u$-direction $J_1$
is much weaker than that along the $x$-direction $J_2$.
Note that $J_2$/$J_1$ is estimated to be 64$^2$
for the (3$x^2$$-$$r^2$)-type FO structure.
Thus, due to the orbital anisotropy,
the spin correlation along the zigzag path is weakened
to suppress the effect of frustration.
On the other hand,
the spin correlation on the rung remains robust in the ladder,
since the orbital shape extends to both the leg-
and rung-directions, as shown in Fig.~\ref{fig-result}(c).

In order to consider how the orbital state is affected
by the spin excitation,
we investigate the orbital shape
in the lowest-energy state with $S_{\rm tot}^z$=1,
where $S_{\rm tot}^{z}$ is the $z$-component of total spin.
In the zigzag chain,
the FO structure with the same orbital shape
as that in the ground state is observed.
Namely, the double-chain nature remains
even in the spin-excited state.
Thus, it is expected that the spin excitation should be described by
the zigzag spin chain with large $J_2$/$J_1$.
For the zigzag spin chain,
it has been found that the spin gap decreases exponentially
with increasing $J_2$/$J_1$ \cite{White-zigzag}.
In Fig.~\ref{fig-result}(d),
the system-size dependence of the spin gap is shown.
We find that the spin gap converges to almost zero
in the thermodynamic limit,
as expected by analogy with the zigzag spin chain.
On the other hand,
the FO structure occurs also in the ladder,
although the orbital shape slightly changes
from that in the ground state.
A finite spin gap is observed in the ladder,
as shown in Fig.~\ref{fig-result}(d).
Details of the spin-orbital structure
in the spin-excited state will be discussed elsewhere.

In summary,
the zigzag chain is effectively decoupled to two chains
by orbital ordering to suppress the effect of frustration.
Accordingly, the spin gap is approximately zero
in the zigzag chain, while a finite spin gap exists in the ladder.

One of the authors (T.H.) has been supported by a Grant-in-Aid from
the Ministry of Education, Culture, Sports, Science and Technology
of Japan.



\end{document}